\documentclass[sigconf]{acmart}
\usepackage[hide]{chato-notes}
\usepackage{algorithm}
\usepackage{algpseudocode}
\usepackage{colortbl}
\usepackage{array}
\usepackage{booktabs}
\usepackage{xcolor}
\usepackage[all]{nowidow}
\usepackage{balance}

\newcommand{\craig}[1]{#1}
\newcommand{\io}[1]{#1}
\newcommand{\sm}[1]{#1}
\newcommand{\smm}[1]{#1}

\newcommand{\midsepremove}{\aboverulesep = 0mm \belowrulesep = 0mm}
\midsepremove

\newcommand{\greyrule}{\arrayrulecolor{black!30}\midrule\arrayrulecolor{black}}

\usepackage{marginnote}
\newcommand{\pageenlarge}[1]{\enlargethispage{#1\baselineskip}}

\setcopyright{acmcopyright}
\copyrightyear{}
\acmYear{}
\acmDOI{}

\acmConference[]{}{}{}
\acmBooktitle{}
\acmPrice{}
\acmISBN{}

\setcopyright{rightsretained}

\begin{document}
\fancyhead{}
\pagestyle{plain}

\title[IntenT5: Search Result Diversification using Causal Language Models]{IntenT5: Search Result Diversification \\ using Causal Language Models}

\author{Sean MacAvaney, Craig Macdonald, Roderick Murray-Smith, Iadh Ounis}
\email{{sean.macavaney,craig.macdonald,roderick.murray-smith,iadh.ounis}@glasgow.ac.uk}
\affiliation{%
  \institution{University of Glasgow}
  \city{Glasgow}
  \country{UK}
}

\settopmatter{printacmref=false}

\renewcommand{\shortauthors}{MacAvaney, et al.}

\begin{abstract}
Search result diversification is a beneficial approach to overcome under-specified queries, such as those that are ambiguous or multi-faceted. Existing approaches often rely on massive query logs and interaction data to generate a variety of possible query intents, which then can be used to re-rank documents. However, relying on user interaction data is problematic because one first needs a massive user base to build a sufficient log; public query logs are insufficient on their own. Given the recent success of causal language models (such as the Text-To-Text Transformer (T5) model) at text generation tasks, we explore the capacity of these models to generate potential query intents. We find that to encourage diversity in the generated queries, it is beneficial to adapt the model by including a new Distributional Causal Language Modeling (DCLM) objective during fine-tuning and a representation replacement during inference. Across six standard evaluation benchmarks, we find that our method (which we call IntenT5) improves search result diversity and attains (and sometimes exceeds) the diversity obtained when using query suggestions based on a proprietary query log. Our analysis shows that our approach is most effective for multi-faceted queries and is able to generalize effectively to queries that were unseen in training data.
\end{abstract}

\maketitle

\pageenlarge{1}\section{Introduction}

\begin{figure}
\centering
\begin{tabular}{|p{0.2cm}p{7.75cm}|}
\hline
\multicolumn{2}{|l|}{\bf (a) monoT5 (no diversity)}\\
1 (A)&Penguins (order Sphenisciformes, family Spheniscidae) are a group of aquatic, flightless birds living almost exclusively in the Southern Hemisphere, especially in Antarctica... \\
2 (A)&Penguins are a group of aquatic, flightless birds. They live almost exclusively in the Southern Hemisphere, with only one species, the Galapagos penguin, found north of the equator... \\
3 (A)&Penguins are flightless birds that are highly adapted for the marine environment. They are excellent swimmers, and can dive to great depths, (emperor penguins can dive to over.. \\
4 (A)&Penguins are an iconic family of aquatic, flightless birds. Although we think of penguins as Antarctic birds some, like the Galapagos penguin, live in much warmer climates near... \\
5 (A)&Penguins are torpedo-shaped, flightless birds that live in the southern regions of the Earth. Though many people imagine a small, black-and-white animal when they think of penguins... \\
\hline
\multicolumn{2}{|l|}{\bf (b) monoT5 (using IntenT5 + DCLM + RS + xQuAD)}\\
1 (A) & Penguins (order Sphenisciformes, family Spheniscidae) are a group of aquatic, flightless birds living almost exclusively in the Southern Hemisphere, especially in Antarctica... \\

2 (H) & Television coverage of the event will be provided by WPXI beginning at 12 p.m. ET and running through the entire event. You can also stream the parade live on the Penguins'... \\

3 (A) & The most abundant species of penguin is the ``Macaroni'' with an approximate population of 20 to 25 Million individuals while there are only around 1,800 Galapagos penguins left...\\

4 (H) & Marc-Andre Fleury will get the start in goal for Pittsburgh on Saturday. The Penguins are in Buffalo for the season finale and need a win to clinch a playoff spot. Fleury was in goal... \\

5 (H) & It is the home of the Pittsburgh Penguins, who moved from their former home across the street prior to the 2010-11 NHL season. The CONSOL Energy Center has 18,387 seats for... \\
\hline
\end{tabular}
\caption{Top results for the query ``penguins'' re-ranked using \io{monoT5} without diversity measures and using our IntenT5 approach on the MS MARCO passage corpus. We manually removed near-duplicate passages from the results. Note that the IntenT5 model has a better variety of top-ranked passages related to both the (A)nimal and (H)ockey.}
\label{fig:penguins}
\end{figure}

\looseness -1 Although contextualized language models (such as BERT~\cite{Devlin2019BERTPO} and T5~\cite{Raffel2020ExploringTL}) have been shown to be highly effective at adhoc ranking~\cite{Nogueira2019PassageRW,macavaney:sigir2019-cedr,Nogueira2020DocumentRW}, they perform best with queries that give adequate context, such as natural-language questions~\cite{Dai2019DeeperTU}. Despite the rise of more expressive querying techniques (such as in conversational search systems~\cite{Radlinski2017ATF}), keyword-based querying remains a popular choice for users.\footnote{\url{https://trends.google.com}} However, keyword queries can often be under-specified, giving rise to multiple possible interpretations or \textit{intents}~\cite{wt09}. Unlike prior lexical models, which do not account for word senses or usage in context, contextualized language models are prone to scoring based on a single predominant sense, which can hinder search result quality for under-specified queries. For instance, the results in Figure~\ref{fig:penguins}(a) \io{are} all similar and do not cover a variety of information needs.

\pageenlarge{1} \looseness -1 Under-specified queries can be considered \textit{ambiguous} and/or \textit{faceted}~\cite{nrbp}. For ambiguous queries, intents are distinct and often correspond to different word senses. For example, the query ``penguins'' may refer to either the animal or the American ice hockey team (among other senses). In Figure~\ref{fig:penguins}, we see that the monoT5 model~\cite{Nogueira2020DocumentRW} only identifies passages for the former sense in the top results. In fact, the first occurrence of a document about the hockey team is ranked at position 158, likely meaning that users with this query intent would likely need to reformulate their query to satisfy their information need. In a faceted query, a user may be interested in different aspects about a given topic. In the example of ``penguins'', a user may be looking for information about the animal's appearance, habitat, life cycle, etc., or the hockey team's schedule, roster, score, etc. Here again, the monoT5 results \io{also} lack diversity in terms of facets, with the top results all focusing on the habitat and appearance.

Search result diversification approaches aim to overcome this issue. In this setting, multiple potential query intents are predicted and the relevance scores for each intent are combined to provide diversity among the top results (e.g., using algorithms like \craig{IA-Select~\cite{agrawal2009diversifying}},  xQuAD~\cite{Santos2010ExploitingQR} or PM2~\cite{Dang2012DiversityBP}). Intents can be inferred from manually-constructed hierarchies~\cite{agrawal2009diversifying} or from interaction data~\cite{Santos2010ExploitingQR}, such as popular searches or reformulations. Although using interaction data is possible for large, established search engines, it is not \io{a feasible approach} for search engines that do not have massive user bases, \craig{nor} to academic researchers, as query logs are \craig{proprietary data}. Researchers have instead largely relied on search result suggestions from major search engines, which are black-box algorithms, or using the ``gold'' intents used for diversity evaluation~\cite{Dang2012DiversityBP}. Thus, an effective approach for generating potential query intents without needing a massive amount of interaction data is desirable.

\begin{figure}
\centering
\includegraphics[scale=0.5]{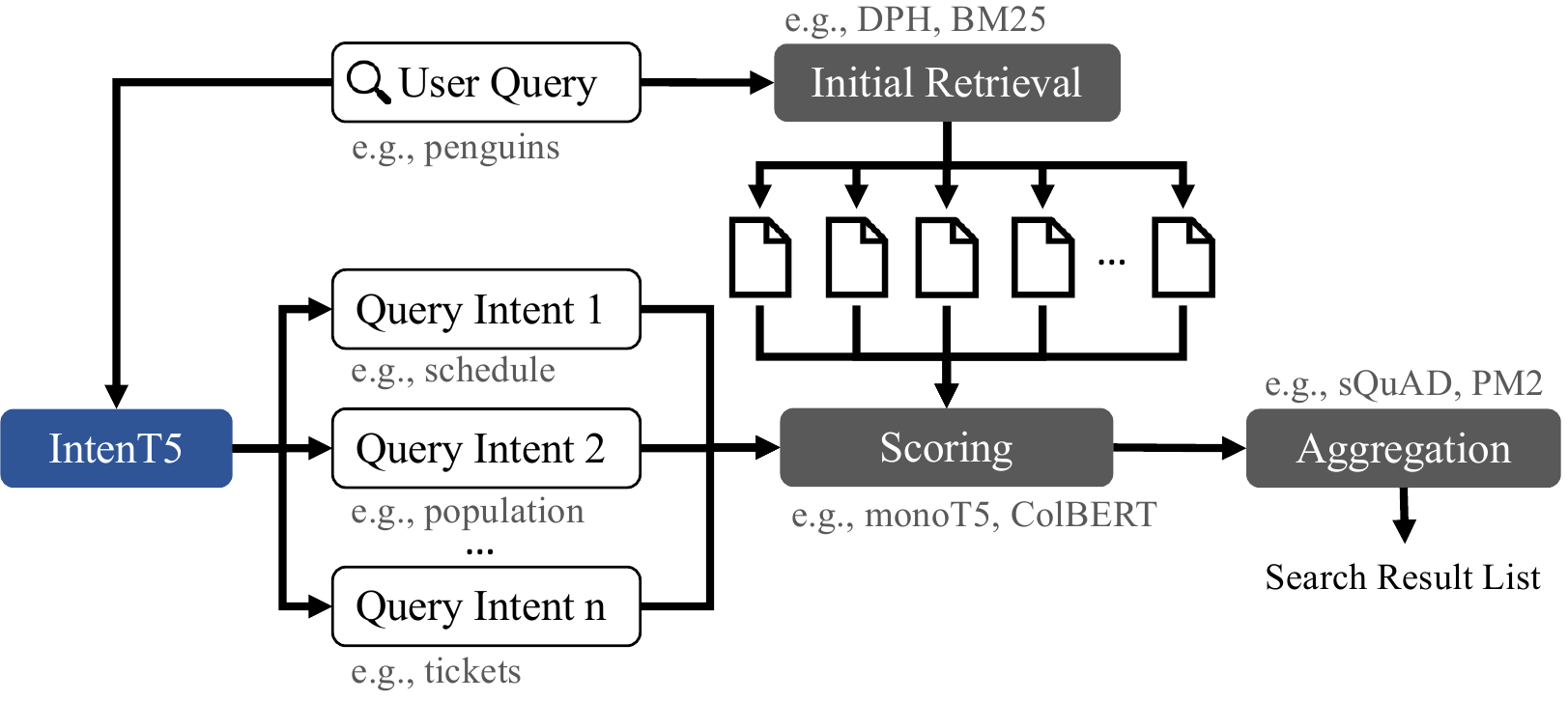}
\vspace{-0.5em}
\caption{Overview of our search result diversification system using IntenT5 to generate potential query intents.}
\label{fig:overview}\vspace{-1em}
\end{figure}

\pageenlarge{2} Given the recent success of Causal Language Models (CLMs) such as T5~\cite{Raffel2020ExploringTL} in a variety of text generation tasks, we propose using these models to generate potential intents for under-specified queries. \io{Figure~\ref{fig:overview} provides an} overview of our approach (IntenT5). We fine-tune the model on a moderately-sized collection of queries (ORCAS~\cite{craswell2020orcas}), and evaluate using 6 TREC diversity benchmark datasets. We find \io{that} our approach improves the search result diversity of both a lexical and neural re-ranking models, and can even exceed the diversity performance when using Google query suggestions and the gold TREC intent descriptions. We also find that our approach has the biggest gains on queries that occur infrequently (or never) in the collection of training queries, showing that the approach is able to generalize effectively to unseen queries.

Through analysis of our proposed IntenT5 model, we find that it has difficulty improving over plain adhoc ranking models for ambiguous queries. Indeed, we find this to be challenging for other approaches as well. In an attempt to better handle ambiguity, we explore two novel techniques for improving the variety of generated intents. First, we propose a \textit{Distributional} Causal Language Modeling (DCLM) objective. This approach targets the observation that a typical CLM trained for this task tends to over-predict general terms that are not specific to the query (e.g., `information', `meaning', `history') \io{since} these are highly-represented in the training data. This approach simultaneously optimizes the model to generate all subsequent tokens that can follow a prefix, rather than just a single term, which should help the model better learn the variety of senses that terms can exhibit. We also introduce a clustering-based \textit{Representation Swapping} (RS) approach that replaces the internal term representations with a variety of possible alternate senses. Qualitatively, we find that these approaches can help improve the diversity of ambiguous queries in isolated cases. For instance, in Figure~\ref{fig:penguins}(b), multiple senses are identified and accounted for. However, in aggreagte, we found insufficient evidence that they improve an unmodified IntenT5 model. Nevertheless, our study opens the door for more research in this area and motivates the creation of larger test sets with ambiguous queries.

In summary, our contributions are:

\begin{itemize}
\item We propose using causal language models for predicting query intents for search result diversification.
\item Across 6 TREC diversity benchmarks, we show that this approach can outperform query intents generated from massive amount of interaction data, and that the model effectively generalizes to previously unseen queries.
\item We introduce a new distributional causal language modeling objective and a representation replacement strategy to better handle ambiguous queries.
\item We provide an analysis that investigates the situations where IntenT5 is effective, and qualitatively assess the generated intents.
\end{itemize}

\pageenlarge{2} \section{Background and Related Work}
In this section, we cover background and prior work related to search result diversification \craig{(Section~\ref{ssec:related:div})}, causal language modeling \craig{(Section~\ref{ssec:related:clm})}, and neural ranking \craig{(Section~\ref{ssec:related:neural})}.

\subsection{Search Result Diversification}\label{ssec:related:div}

Search result diversification techniques aim to handle ambiguous queries. \craig{Early works aimed to ensure that the retrieved documents addressed distinct topics - for instance, Maximal Marginal Relevance (MMR)~\cite{mmr} can be used to promote documents that are relevant to the user's query, but are dissimilar to those document already retrieved. In doing so, the typical conventional document independence assumption inherent \io{to} the Probability Ranking Principle is relaxed. Indeed, by diversifying the topics covered in the top-ranked documents, diversification approaches aim to address the risk that there are no relevant documents retrieved for the user's information need~\cite{10.1145/1571941.1571963}. Other approaches such as IA-Select used category hierarchies to identify documents with different intents~\cite{agrawal2009diversifying}.} 

\craig{Given an under-specified (ambiguous or multi-faceted) query $q$ and a candidate set of documents $d$, \io{the} potential query intents $\{i_1,i_2,..i_j\}$ can be identified as {\em sub-queries}, i.e., query formulations that more clearly identify relevant documents \io{about a particular interpretation of the query}.} 
These intents are usually identified from interaction data, such as query logs~\cite{Santos2010ExploitingQR}. In academic research, query suggestions from major search engines often serve as a stand-in for this process~\cite{Dang2012DiversityBP,Santos2010ExploitingQR}. The candidate documents are re-scored for each of the intents. The scores from the individual intents are then aggregated into a final re-ranking of documents, using an algorithm such as xQuAD or PM2.

\looseness -1 Aggregation strategies \io{typically} attempt to balance relevance and novelty. xQuAD~\cite{Santos2010ExploitingQR} iteratively selects documents that exhibit high relevance to the original query and are maximally relevant to the set of intents. As documents are selected, the relevance scores of documents to intents already shown are marginalized. The balance between relevance to the original query and the intents are controlled with a parameter $\lambda$. PM2~\cite{Dang2012DiversityBP} is an aggregation strategy based \io{on} a proportional representation voting scheme. This aggregation strategy ignores relevance to the original query, and iteratively selects the intent least represented so far. The impact of the selected intent and the intents that are not selected when choosing the next document is controlled with a parameter $\lambda$ (not to be confused with xQuAD's $\lambda$).

\craig{More recently, there has been a family of work addressing learned models for diversification~\cite{10.1145/3209978.3209979, 10.1145/3077136.3080775,YIGITSERT2020102356}; we see this as orthogonal to our work here, \io{since we} do not consider learned diversification approaches.} Indeed, in this work, we study the process of query intent generation for search result diversification, rather than aggregation strategies. For further information about search result diversification, see~\cite{Santos2015SearchRD}.

\subsection{Causal Language Modeling}\label{ssec:related:clm}

\pageenlarge{1}  \looseness -1 Causal Language Models (CLMs) predict the probability of a token $w_k$ given the prior tokens in the sequence: $P(w_k|w_{k-1},w_{k-2},...,w_{1})$. This property makes a CLM able to generate text: by providing a prompt, the model can iteratively predict a likely sequence of following tokens. However, a complete search of the space is exponential because the probability of each token depends on the preceding generated tokens. Various strategies exist for pruning this space. A popular approach for reducing the search space is a beam search, where a fixed number of high-probability sequences are explored in parallel. Alternative formulations, such as Diverse Beam Search~\cite{Vijayakumar2016DiverseBS} have been proposed, but we found these techniques unnecessary for short texts like queries. We refer the reader to \citet{Meister2020IfBS} for more further details about beam search and text generation strategies.

While CLMs previously accomplished modeling with recurrent neural networks~\cite{Mikolov2010RecurrentNN,Jzefowicz2016ExploringTL}, this modeling has recently been accomplished through transformer networks~\cite{Dai2019TransformerXLAL}. Networks pre-trained with a causal language modeling objective, for instance T5~\cite{Raffel2020ExploringTL}, can be an effective starting point for further task-specific training~\cite{Raffel2020ExploringTL}. \sm{In the case of T5, specific tasks are also cast as sequence generation problems by encoding the source text and generating the model prediction (e.g., a label for classification tasks).}

In this work, we explore the capacity of CLMs (T5 in particular) for generating a diverse set of possible query intents. This differs from common uses of CLMs due to the short nature of the text (keyword queries rather than natural-language sentences or paragraphs) and the focus on the diversity of the predictions.

\subsection{Neural Ranking}\label{ssec:related:neural}

\looseness -1 Neural approaches have been shown to be effective for adhoc ranking tasks, especially when the intents are expressed clearly and in natural language~\cite{Dai2019DeeperTU}. Pre-trained contextualized language models, such as BERT~\cite{Devlin2019BERTPO} and ELECTRA~\cite{Clark2020ELECTRAPT} have been particularly effective for adhoc ranking~\cite{Nogueira2019PassageRW,macavaney:sigir2019-cedr}. A simple application of these models is the ``vanilla'' setting (also called CLS, mono, and cross), where the query and document text is jointly encoded and the model's classification component is tuned to provide a ranking score. Due to the expense of such approaches, neural models are typically used as re-rankers; that is, an initial ranking model such as BM25 is used to provide a pool of documents that can be re-ranked by the neural method. Neural approaches have also been applied as first-stage rankers~\cite{Zamani2018FromNR,Khattab2020ColBERTEA,Xiong2020ApproximateNN} \craig{(also called dense retrieval)}. The ColBERT \io{model}~\cite{Khattab2020ColBERTEA} scores documents \craig{based on} BERT-based query and document term representations. The similarity between the query and document representations are summed to give a ranking score. This model can be used as both a first-stage dense ranker (through an approximate search over its representations) as well as a re-ranker (to produce precise ranking scores). We focus on re-ranking approaches in this work, which is in line with prior work on diversification~\cite{Santos2015SearchRD}.

CLMs have been used in neural ranking tasks. \citeauthor{Nogueira2020DocumentRW}~\cite{Nogueira2020DocumentRW} \craig{predicted} the relevance of a document to a query using the T5 model (monoT5). \citet{Pradeep2021TheED} further \io{explored} this model and \io{showed} that it can be used in conjunction with a version that scores and aggregates pairs of documents (duoT5). Both these models only use CLMs insofar as predicting a single token (`true' or `false' for relevance) given the text of the query and document and a prompt. Here, the probability of `true' is used as the ranking score. Doc2query models~\cite{Nogueira2019DocumentEB,nogueiradoc2query} generate possible queries, conditioned on document text, to include in an inverted index.

\looseness -1 Unlike prior neural ranking efforts, we focus on diversity ranking, rather than adhoc ranking. Specifically, we use a neural CLM to generate possible query intents given \io{a} query text, which differs from prior uses of CLMs in neural ranking. These query intents are then used to score and re-rank documents. To our knowledge, this is the first usage of neural ranking models for diversity ranking. For further details on neural ranking and re-ranking models, see~\cite{Mitra2018AnIT,Lin2021PretrainedTF}.

\pageenlarge{1} \section{Generating Query Intents}
In this section, we describe our proposed IntenT5 model for query intent generation (Section~\ref{ssec:meth:intent}), as well as two model adaptations \craig{intended} to improve the handling of ambiguous queries: distributional causal language modeling (Section~\ref{ssec:meth:dclm}) and representation swapping (Section~\ref{ssec:meth:rs}).

\subsection{IntenT5}\label{ssec:meth:intent}

Recall that we seek to train a model that can be used to generate potential query intents. We formulate this task as a sequence-to-sequence generation problem by predicting additional terms for the user's initial (under-specified) query.

We first fine-tune a T5~\cite{Raffel2020ExploringTL} model using a causal language modeling objective over a collection of queries. Note that this training approach does not require search frequency, session, or click information; it only requires a collection of query text. This makes a variety of data sources available for training, such as the ORCAS~\cite{craswell2020orcas} \craig{query} collection. This is a desirable quality because releasing query text poses fewer risks to personal privacy than more extensive interaction information.

Recall that for a sequence $w$ consisting of $k$ tokens, causal language models optimize for $P(w_k|w_{k-1},w_{k-2},...,w_{1})$.
To generate intents, we use a beam search to identify highly-probable sequences. No length penalization is applied, but queries are limited to 10 generated tokens.\footnote{We found that this covers the vast majority of cases for this task, in practice.} We apply basic filtering techniques to remove generated intents that do not provide adequate additional context. In particular, we first remove terms that appear in the original query. Since neural retrieval models are sensitive to word morphology~\cite{macavaney:arxiv2020-abnirml}, we only consider exact term matches for this filter. We also discard intents that are very short (less than 6 characters, e.g., ``.com''), as we found that these usually carry little valuable context. Among the filtered intents, we select the top $n$ most probable sequences. Note that this generation process is fully deterministic, so the results are \craig{entirely} replicable.

Each retrieved document is scored for each of the generated intents, and the intent scores are aggregated using an established \craig{diversification} algorithm, like xQuAD~\cite{Santos2010ExploitingQR} or PM2~\cite{Dang2012DiversityBP}.

\io{Instead of T5}, our approach could also be applied to other pre-trained causal language models, such as BART~\cite{lewis-etal-2020-bart}, \craig{\io{however}, we leave \io{such a study}} for future work.

\begin{figure}
\centering
\begin{tabular}{|p{4cm}p{4cm}|}
\hline
\multicolumn{2}{|l|}{\textbf{Documents} (MS MARCO passages)} \\
\multicolumn{2}{|l|}{Penguins are a group of aquatic, flightless birds. They live...} \\
\multicolumn{2}{|l|}{Penguins are birds, not mammals. Many bird species are...} \\
\multicolumn{2}{|l|}{Penguins are carnivores with piscivorous diets, getting all...} \\
\multicolumn{2}{|l|}{Penguins are especially abundant on islands in colder climates...} \\
\hline
\textbf{Keyword Queries} (ORCAS) & penguins hockey \\
penguins adaptations & penguins hockey game \\
penguins animals & penguins hockey game tonight \\
penguins animals facts & penguins hockey live streaming \\
\hline
\end{tabular}
\caption{Comparison of document language and query language. Queries naturally lend themselves to a tree structure, motivating our DCLM approach.}\vspace{-\baselineskip}
\label{fig:doc_query_lang}
\end{figure}

\pageenlarge{1} \subsection{Distributional Causal Language Modeling}\label{ssec:meth:dclm}

Typical natural language prose, such as the type of text found in documents, lends itself well to CLM because the text quickly diverges into a multitude of meanings. For instance, in Figure~\ref{fig:doc_query_lang}, we see that the prefix ``Penguins are'' diverges into a variety of sequences \sm{(e.g., ``a group of'', ``birds, not'', ``carnivores with'', etc.)}. If structured by prefixes, this results in long chains of tokens. Keyword queries, on the other hand, typically have a hierarchical prefix-based nature. When structured as a tree, it tends to be shallow and dense. For instance, in Figure~\ref{fig:doc_query_lang}, a distribution of terms is likely to follow the prefix `penguins' (e.g., adaptations, animals, hockey, etc.). Similarly, a distribution of terms follows the prefix `penguins hockey' (e.g., game, live, news, score, etc.).

\begin{figure}
\centering \vspace{-\baselineskip}
\includegraphics[scale=0.7]{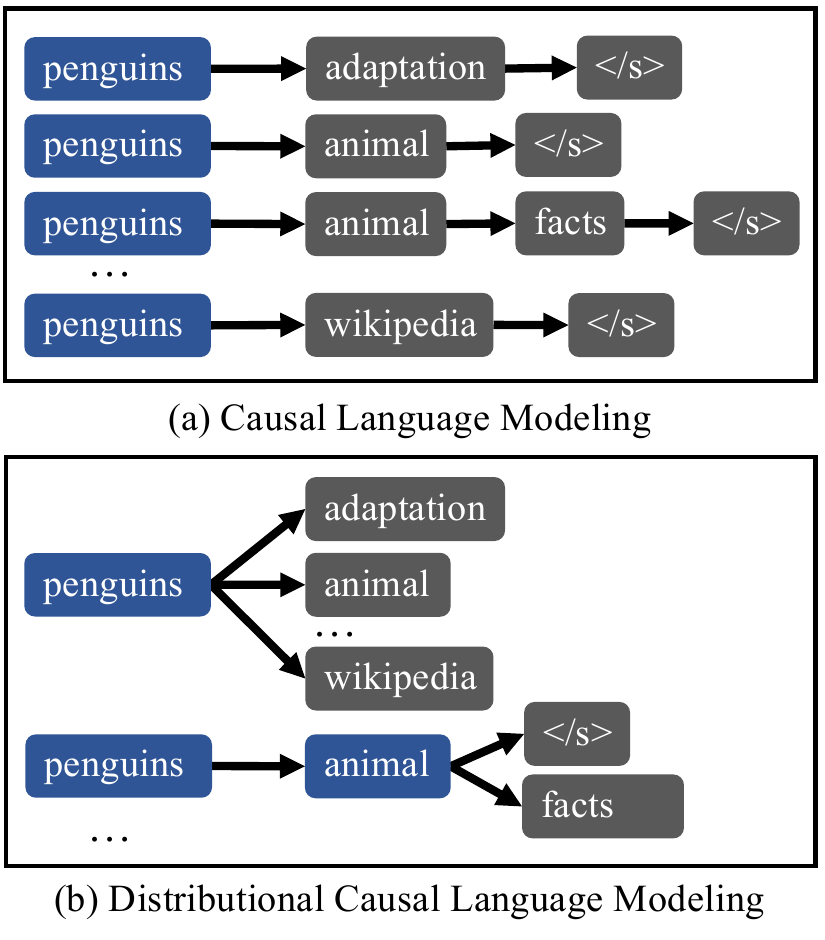}
\vspace{-1em}
\caption{Graphical distinction between Causal Language Modeling (CLM, (a)) and our proposed Distributional Causal Language Modeling (DCLM, (b)). Given a prompt (e.g., `penguins'), a DCLM objective optimizes for all possible subsequent tokens (e.g., \textit{adaptation}, \textit{animal}, \textit{antarctica}, etc.) rather than implicitly learning this distribution over numerous training samples.}\vspace{-\baselineskip}
\label{fig:dclm}
\end{figure}

Based on this observation, we propose a new variant of Causal Language Modeling (CLM) designed for keyword queries: Distributional Causal Language Modeling (DCLM). In contrast with CLM, DCLM considers other texts in the source collection when building the learning objectives through the construction of a prefix tree. In other words, while CLM considers each sequence independently, DCLM builds a distribution of terms that follow a given prefix. Visually, the difference between the approaches are shown in Figure~\ref{fig:dclm}. When training, the prefix tree is used to find all subsequent tokens across the collection given a prefix and optimizes the output of the model to generate \textit{all} these tokens (with equal probability). The training process for DCLM is given in Algorithm~\ref{alg:dclm}.

\begin{algorithm}[tb]
\caption{DCLM Training Procedure}
\label{alg:dclm}
\begin{algorithmic}
\State $\text{tree} \gets \text{BuildPrefixTree}(\text{corpus})$
\Repeat
    \State $\text{prefix} \gets \{\text{RandomSelect(tree.children)}\}$
    \State $\text{targets} \gets \text{prefix.children}$
    \While{$|\text{prefix}[-1]\text{.children}|>0$}
        \State Optimize $P(\text{prefix}[-1]\text{.children}|\text{prefix})$
        \State $\text{prefix} = \{\text{prefix}, \text{RandomSelect}(\text{prefix}[-1]\text{.children})\}$
    \EndWhile
\Until{converged}
\end{algorithmic}
\end{algorithm}

\pageenlarge{1} \subsection{Representation Swapping}\label{ssec:meth:rs}

\sm{In a transformer model -- \craig{which is} the underlying neural architecture of T5 -- tokens are represented as contextualized vector representations. These representations map tokens to a particular sense -- \craig{this is} exploited by several neural ranking models (like CEDR~\cite{macavaney:sigir2019-cedr}, TK~\cite{Hofsttter2020InterpretableT}, and ColBERT~\cite{Khattab2020ColBERTEA}) to match particular word senses. Normally, the surrounding words in a piece of text offer adequate context to disambiguate word senses. However, short queries inherently lack such context. For instance, in the case where a query contains only a single term, we find that transformer models simply choose a predominant sense (e.g., the animal sense for the query \textit{penguins}). When used with the IntenT5 model, we find that this causes \craig{the} generated intents to lack diversity of word senses (\craig{we demonstrate this in} Section~\ref{sec:analysis}). We introduce an approach we call Representation Swapping (RS) to overcome this issue.}

\sm{
RS starts by building a set of $k$ prototype representations for each term in a corpus.\footnote{In practice, terms are filtered to only those meeting a frequency threshold, as infrequent terms are less prone to being ambiguous. In an offline experimentation setting, only \craig{the} terms that appear in \craig{the} test queries need to be considered.} For a given term, a random sample of passages from a corpus that contain \craig{the} term are selected. Then, the term's internal representations are extracted for each passage.\footnote{\sm{Note that terms are often broken down into subwords by T5's tokenizer. In this case, we concatenate the representations of each of the constituent subwords. Although the size of the concatenated representations across terms may differ, the representations for a single term are the same length.}} Note that because of the context from other terms in the passage, these representations include the sense information. \craig{All of the these representations for a given term} are then clustered into $k$ clusters. A single prototype representation for each cluster is selected by finding the representation \io{that is} closest to the median value across \io{the representations} in the cluster. Using this approach, we find that the sentences from which the prototypes were selected often express different word senses.}

\sm{
\looseness -1 Finally, when processing a query, the IntenT5 model is executed multiple times: once with the original representation (obtained from encoding the query alone), and then $k$ additional times for each term. In these instances, the internal representations of a given term are swapped with the prototype representation. This allows the T5 model to essentially inherit the context from the prototype sentence for the ambiguous query, and allows the model to generate text based on different senses. This approach only needs to be applied \craig{for} shorter queries, \io{since} longer queries provide enough context on their own. This introduces a parameter $l$, as the maximum query length. The final intents generated by the model are selected using a diversity aggregation algorithm like xQuAD,
which ensures a good mix of possible senses in the generated intents.}

\pageenlarge{1}\section{Experimental Setup}
\looseness -1 We experiment to answer the following research questions:

\textbf{RQ1}: Can the intents generated from IntenT5 be used to diversify search results?

\textbf{RQ2}: Do queries that appeared in the IntenT5 training data perform better than those that did not?

\textbf{RQ3}: Does IntenT5 perform better at ambiguous or multi-faceted queries?

\textbf{RQ4}: Does training with a distributional causal language modeling objective or performing representation swapping improve the quality of the generated intents?

\subsection{IntenT5 Training and Settings}

Although IntenT5 can be trained on any moderately large collection of queries, we train the model using the queries from the ORCAS~\cite{craswell2020orcas} dataset. With 10.4M unique queries, this dataset is both moderately large\footnote{It is widely known that Google processes over 3B queries daily, with roughly 15\% being completely unique.} and easily accessible to researchers without signed data usage agreements. The queries in ORCAS were harvested from the Bing logs, filtered down to only those where users clicked on a document found in the MS MARCO~\cite{Campos2016MSMA} document dataset. Queries in this collection contain an average of 3.3 terms, with the majority of queries containing either 2 or 3 terms. \smm{The IntenT5 model is fine-tuned from \texttt{t5-base} using default parameters (learning rate: $5\times10^{-5}$, 3 epochs, Adam optimizer).}

\sm{When applying RS, we use $k=5$, $l=1$, xQuAD with $\lambda=1$, and use agglomerative clustering, based on qualitative observations during pilot studies. We select 1,000 passages per term from the MS MARCO document corpus, \craig{so} as not to potentially bias our results to our test corpora.}

\subsection{Evaluation}

\begin{table}
\centering
\caption{Dataset statistics.}
\label{tab:datasets}
\begin{tabular}{lrrrr}
\toprule
        &           & \# Terms  & \# Subtopics & \# Pos. Qrels \\
Dataset & \# Topics & per topic & (per topic)  & (per topic) \\
\midrule
WT09~\cite{wt09} & 50 & 2.1 & 243 (4.9) &  6,499 (130.0) \\
WT10~\cite{wt10} & 50 & 2.1 & 218 (4.4) &  9,006 (180.1) \\
WT11~\cite{wt11} & 50 & 3.4 & 168 (3.4) &  8,378 (167.6) \\
WT12~\cite{wt12} & 50 & 2.3 & 195 (3.9) &  9,368 (187.4) \\
WT13~\cite{wt13} & 50 & 3.3 & 134 (2.7) &  9,121 (182.4) \\
WT14~\cite{wt14} & 50 & 3.3 & 132 (2.6)  &10,629 (212.6) \\
\midrule
Total            &300 & 2.8 &1,090 (3.6)& 53,001 (176.7) \\
\bottomrule
\end{tabular}
\end{table}

We evaluate the effectiveness of our approach on the TREC Web Track (WT) 2009--14 diversity benchmarks~\cite{wt09,wt10,wt11,wt12,wt13}, consisting of 300 topics and 1,090 sub-topics in total.  Table~\ref{tab:datasets} \io{provides a} summary of these datasets. These benchmarks span two corpora: WT09--12 use ClueWeb09-B (50M documents) and WT13--14 use ClueWeb12-B13 (52M documents). We use the keyword-based ``title'' queries, simulating a setting where the information need is under-specified. To the best of our knowledge, these are the largest and most extensive public benchmarks for evaluating search result diversification.

\looseness -1 \pageenlarge{1} We measure system performance with three diversification-aware variants of standard evaluation measures: $\alpha$-nDCG@20~\cite{alphandcg}, ERR-IA@20~\cite{erria}, and NRPB~\cite{nrbp}. These are the official task evaluation metrics for WT10--14 (WT09 used $\alpha$-nDCG@20 and P-IA@20). $\alpha$-nDCG is a variant of nDCG~\cite{ndcg} that accounts for the novelty of topics introduced. We use the default $\alpha$ parameter (the probability of an incorrect positive judgment) of 0.5 for this measure. ERR-IA is a simple mean over the Expected Reciprocal Rank of each intent, since WT09--14 weight the gold query intents uniformly. NRBP (Novelty- and Rank-Biased Precision) is an extension of RBP~\cite{Moffat2008RankbiasedPF}, measuring \io{the average} utility gained as a user scans the search results. Metrics were calculated from the official task evaluation script \texttt{ndeval}.\footnote{\url{https://trec.nist.gov/data/web/10/ndeval.c}}  \craig{Furthermore, as these test collections were created before the advent of neural ranking models,} we also report the judgment rate among the top 20 results (Judged@20) \craig{to ascertain the completeness of the relevance assessment pool in the presence of such neural \io{models}.}

To test the significance of differences, we use paired t-tests with $p<0.05$, accounting \io{for} multiple tests where appropriate with a Bonferroni correction. In some cases, we test for significant equivalences (i.e., that the means are the same). For these tests, we use a two one-sided test (TOST~\cite{walker2011understanding}) with $p<0.05$. Following prior work using \craig{a} TOST for retrieval effectiveness~\cite{Mackenzie2018QueryDA}, we set the acceptable equivalence range to $\pm0.01$.

\subsection{Baselines}

\looseness -1 To \craig{put} the performance of our method in context, we include several adhoc and diversity baselines. As adhoc baselines, we compare with DPH~\cite{Amati2008FUBIA}, a lexical model (we found it to perform better than BM25 in pilot studies), Vanilla BERT~\cite{macavaney:sigir2019-cedr}, monoT5~\cite{Nogueira2020DocumentRW}, and a ColBERT~\cite{Khattab2020ColBERTEA} re-ranker.\footnote{We only use ColBERT in a re-ranking setting due to the challenge \craig{in scaling its space intensive dense retrieval indices to the very large} ClueWeb datasets.} Since the neural models we use have a maximum sequence length, we apply the MaxPassage~\cite{Dai2019DeeperTU} scoring approach. Passages are constructed using sliding windows of 150 tokens (stride 75). For Vanilla BERT, we trained a model on MS MARCO using the original authors' released code. For monoT5 and ColBERT, we use versions released by the original authors that were trained on the MS MARCO \io{dataset}~\cite{Campos2016MSMA}. This type of zero-shot transfer from MS MARCO to other datasets has generally shown to be effective~\cite{Nogueira2020DocumentRW,macavaney:emnlp2020-sledge}, and reduces the risk of over-fitting to the test collection.

\textbf{Google Suggestions}. We compare with the search suggestions provided by the Google search engine through their public API. Though the precise details of the system are private, public information states that interaction data plays a big role in the \craig{generation of their search suggestions}~\cite{sullivan_2018}, meaning \io{that} this is a strong baseline for an approach based on a query log. Furthermore, this technique was used \craig{extensively} in prior search result diversification work as a source of query intents~\cite{Dang2012DiversityBP,Santos2010ExploitingQR}. Note that the suggestions are sensitive to language, geographic location and current trends; we use the suggestions in English for United States (since the TREC assessors were based in the US); \sm{we will release a copy of these suggestions for reproducibility.}

\pageenlarge{1} \looseness -1 \textbf{Gold Intents}. We also compare with systems that use the ``Gold'' intents provided by the TREC task. Note that this is not a realistic system, as these intents represent the evaluation criteria and \craig{are not known} a priori. \io{Furthermore}, the text of these intents are provided in natural language (similar to TREC description queries), unlike the keyword-based queries they elaborate upon. \craig{Hence, these Gold intents are often reported to represent a potential upperbound on diversification effectiveness, however, later we will show that intents generated by IntenT5 can actually outperform these Gold intents.\footnote{\craig{For instance, this may be caused by an intent identified by an IntenT5 model resulting in a more effective ranking than the corresponding Gold intent.}}} %

\subsection{Model Variants and Parameter Tuning}

We aggregate the intents from IntenT5, Google suggestions, and the Gold intents using xQuAD~\cite{Santos2010ExploitingQR} and PM2~\cite{Dang2012DiversityBP}, representing two strong unsupervised aggregation techniques.\footnote{\looseness -1 Although ``learning-to-diversify'' approaches, such as M$^2$Div~\cite{Feng2018FromGS}, can be effective, our work focuses on \textit{explicit} diversification. The explicit setting is advantageous because it allows for a greater degree of model interpretability and transparency with the user.} For all these models, we tune the number of generated intents and the aggregation $\lambda$ parameter using a grid search over the remaining collections (e.g., WT09 parameters are tuned in WT10--14). We search between 1--20 intents (step 1) and $\lambda$ between 0--1 (step 0.1). Neural models re-rank the top 100 DPH results. In summary, an initial pool of 100 documents is retrieved using DPH. Intents are then chosen using IntenT5 (or \io{using the} baseline methods). The documents are then re-scored for each intent using DPH, Vanilla BERT, monoT5, or ColBERT. The scores \io{are} then aggregated using xQuAD or PM2.

\section{Results}\label{sec:res}

\craig{In this section, we provide results for research questions concerning the overall effectiveness of IntentT5 (Section~\ref{ssec:res:rq1}), the impact of queries appearing in the training data (Section~\ref{ssec:res:rq2}), the types of under-specification (Section~\ref{ssec:res:rq3}) and finally the impact on ambiguous queries in particular (Section~\ref{ssec:res:rq4}). Later in Section~
\ref{sec:analysis}, we provide a qualitative analysis of the generated queries}.

\begin{table}
\centering
\caption{Results over the combined TREC WebTrack 2009--14 diversity benchmark datasets. $\alpha$-nDCG, ERR-IA, and Judged are computed with a rank cutoff of 20. The highest value in each section is listed in bold. PM and xQ indicate the PM2 and xQuAD aggregators, respectively. Statistically significant differences between each value and the corresponding non-diversified baseline are indicated by~* (paired t-test, Bonferroni correction, $p<0.05$).}
\label{tab:main_results}
\begin{tabular}{lcrrrr}
\toprule
System & Agg. & $\alpha$-nDCG & ERR-IA & NRBP & Judged \\
\midrule
DPH            &    &   0.3969 &   0.3078 &   0.2690 & 62\% \\
\greyrule
 + IntenT5     & PM & * 0.4213 & * 0.3400 & * 0.3053 & 56\% \\
 + Google Sug. & PM & * 0.4232 & * 0.3360 & * 0.3000 & 58\% \\
 + Gold        & PM & \bf * 0.4566 & \bf * 0.3629 & * \bf 0.3254 & 53\% \\
\greyrule
 + IntenT5     & xQ &   0.4049 &   0.3213 &   0.2845 & 58\% \\
 + Google Sug. & xQ & * 0.4203 & * 0.3326 &   0.2963 & 59\% \\
 + Gold        & xQ & \bf * 0.4545 & \bf * 0.3616 & \bf * 0.3230 & 56\% \\
\midrule
Vanilla BERT   &    &   0.3790 &   0.2824 &   0.2399 & 54\% \\
\greyrule
 + IntenT5     & PM & \bf * 0.4364 & \bf * 0.3475 & \bf * 0.3104 & 60\% \\
 + Google Sug. & PM & * 0.4110 & * 0.3119 &   0.2689 & 57\% \\
 + Gold        & PM & * 0.4214 & * 0.3214 & * 0.2803 & 50\% \\
\greyrule
 + IntenT5     & xQ & \bf * 0.4328 & \bf * 0.3448 & \bf * 0.3085 & 59\% \\
 + Google Sug. & xQ & * 0.4120 & * 0.3140 & * 0.2722 & 59\% \\
 + Gold        & xQ & * 0.4228 & * 0.3236 & * 0.2813 & 55\% \\
\midrule
monoT5            &    &   0.4271 &   0.3342 &   0.2943 & 58\% \\
\greyrule
 + IntenT5        & PM & * 0.4510 &   0.3589 &   0.3213 & 58\% \\
 + Google Sug.    & PM & * 0.4506 &   0.3567 &   0.3181 & 58\% \\
 + Gold           & PM & \bf * 0.4722 & \bf * 0.3777 & \bf * 0.3409 & 58\% \\
\greyrule
 + IntenT5        & xQ &   0.4444 &   0.3549 &   0.3183 & 58\% \\
 + Google Sug.    & xQ & * 0.4492 & * 0.3625 & * 0.3276 & 58\% \\
 + Gold           & xQ & \bf * 0.4574 & \bf * 0.3696 & \bf * 0.3353 & 58\% \\
\midrule
ColBERT        &    &   0.4271 &   0.3334 &   0.2953 & 57\% \\
\greyrule
 + IntenT5     & PM & \bf * 0.4711 & \bf * 0.3914 & \bf * 0.3616 & 58\% \\
 + Google Sug. & PM & * 0.4561 &   0.3654 &   0.3316 & 57\% \\
 + Gold        & PM &   0.4548 &   0.3520 &   0.3106 & 51\% \\
\greyrule
 + IntenT5     & xQ & \bf * 0.4707 & \bf * 0.3890 & \bf * 0.3584 & 63\% \\
 + Google Sug. & xQ & * 0.4552 & * 0.3638 & * 0.3287 & 61\% \\
 + Gold        & xQ & * 0.4560 &   0.3608 &   0.3226 & 56\% \\
\bottomrule
\end{tabular}
\end{table}

\pageenlarge{1}\subsection{RQ1: IntenT5 Effectiveness}\label{ssec:res:rq1}

We present the diversification results for WT09--14 in Table~\ref{tab:main_results}. We generally find that our IntenT5 approach can improve the search result diversity for both lexical and neural models, when aggregating using either PM2 or xQuAD. In fact, there is only one setting (monoT5 scoring with xQuAD aggregation) where diversity is not significantly improved when using IntenT5. The overall best-performing result uses IntenT5 (ColBERT scoring \io{with} PM2 aggregation). These results \craig{also significantly outperform} the corresponding \craig{versions that use} Google suggestions and the Gold intents. Similarly, \craig{when using Vanilla BERT, IntenT5} also significantly outperforms the model using Google suggestions. %
For DPH and monoT5, the diversity effectiveness of IntenT5 is similar to that of the Google suggestions; the differences are not statistically significant. However, through equivalence testing \craig{using a TOST we find that} there is insufficient evidence that the means are equivalent across all evaluation metrics. Curiously, both BERT-based models (Vanilla BERT and ColBERT) are more receptive to the IntenT5 queries than Google suggestions or Gold intents. \smm{This suggests that some underlying language models (here, BERT), may benefit more from artificially-generated intents than others.}

These results provide a clear answer to RQ1: the intents generated from IntenT5 can be used to significantly improve the diversity of search results. Further, they can \craig{also, surprisingly,} outperform Google suggestions and Gold intents.

\begin{table}
\centering
\caption{Diversification performance stratified by the frequency of the query text in ORCAS. Ranges were selected to best approximate 3 even buckets. Statistically significant differences between the unmodified system are indicated by~* (paired t-test, Bonferroni correction, $p<0.05$).}
\label{tab:orcasfreq}
\begin{tabular}{lcrrr}
\toprule
&& \multicolumn{3}{c}{$\alpha$-nDCG@20} \\
\cmidrule(lr){3-5}
System         &Agg.& 0--1     & 2--37    & 38+      \\
\midrule
DPH            &    &   0.4930 &   0.3876 &   0.3048 \\
\greyrule
 + IntenT5     & PM &   0.5217 & * 0.4283 &   0.3091 \\
 + Google Sug. & PM &   0.5026 & * 0.4435 &   0.3204 \\
 + Gold        & PM & \bf * 0.5311 & \bf * 0.4648 & \bf * 0.3704 \\
\greyrule
 + IntenT5     & xQ &   0.4970 & * 0.4177 &   0.2960 \\
 + Google Sug. & xQ &   0.5004 & * 0.4379 &   0.3192 \\
 + Gold        & xQ & \bf * 0.5259 & \bf * 0.4708 & \bf * 0.3640 \\
\midrule
Vanilla BERT   &    &   0.4467 &   0.3756 &   0.3112 \\
\greyrule
 + IntenT5     & PM & \bf * 0.5043 & \bf * 0.4403 & \bf   0.3615 \\
 + Google Sug. & PM &   0.4634 &   0.4186 &   0.3487 \\
 + Gold        & PM & * 0.4831 & * 0.4290 &   0.3492 \\
\greyrule
 + IntenT5     & xQ & \bf * 0.5041 & * 0.4308 & \bf 0.3598 \\
 + Google Sug. & xQ & * 0.4853 &   0.4027 &   0.3439 \\
 + Gold        & xQ & * 0.4770 & \bf * 0.4362 &   0.3531 \\
\midrule
monoT5         &    &   0.5080 &   0.4331 &   0.3364 \\
\greyrule
 + IntenT5     & PM &   0.5262 & * 0.4707 &   0.3530 \\
 + Google Sug. & PM &   0.5305 &   0.4598 &   0.3577 \\
 + Gold        & PM & \bf * 0.5421 & \bf  0.4712 & \bf * 0.3997 \\
\greyrule
 + IntenT5     & xQ &   0.5180 & * 0.4648 &   0.3474 \\
 + Google Sug. & xQ &   0.5217 &   0.4509 & * 0.3714 \\
 + Gold        & xQ & \bf  0.5270 & \bf  0.4657 & \bf * 0.3762 \\
\midrule
ColBERT        &    &   0.4938 &   0.4241 &   0.3600 \\
\greyrule
 + IntenT5     & PM & \bf * 0.5484 & \bf * 0.4934 &   0.3685 \\
 + Google Sug. & PM &   0.5067 &   0.4625 & \bf   0.3968 \\
 + Gold        & PM &   0.5254 &   0.4726 &   0.3635 \\
\greyrule
 + IntenT5     & xQ & \bf * 0.5440 & \bf * 0.4868 &   0.3783 \\
 + Google Sug. & xQ &   0.5141 & * 0.4631 & \bf   0.3857 \\
 + Gold        & xQ &   0.5121 & * 0.4874 &   0.3669 \\
\midrule
(Count)         &    & (105)    & (96)     & (99)     \\
\bottomrule
\end{tabular}
\end{table}

\subsection{RQ2: Effect of Queries in Training Data}\label{ssec:res:rq2}

It is possible that the IntenT5 model simply memorizes the data that is present in the training dataset, rather than utilizing the language characteristics learned in the pre-training process to generalize to new queries. To investigate this, we stratify the dataset into three roughly equal-sized buckets representing the frequency that the query appears in ORCAS. We use simple \craig{case-insensitive} string matching, and count matches if they appear anywhere in the text (not just at the start of the text). We find that roughly one third of \io{the} WebTrack diversity queries either do not appear at all in ORCAS, or only appear once. For these queries, IntenT5 is forced to generalize. The next bucket (2--37 occurrences in ORCAS) contains roughly the next third of \io{the} WebTrack queries, and 38 or more queries forms the final bucket. We present the results of this experiment in Table~\ref{tab:orcasfreq}. Here, we find that our IntenT5 model excels at cases where it either needs to generalize (first bucket) or where memorization is manageable (second bucket); in 11 of the 16 cases, IntenT5 scores higher than the Google suggestions. Further, IntenT5 boasts the overall highest effectiveness in both buckets: 0.5484 and 0.4934, respectively (for ColBERT + IntenT5 + PM2). In the final case, where there are numerous occurrences in the training data, IntenT5 \smm{never significantly} outperforms the baseline system. Unsurprisingly, Google suggestions score higher than IntenT5 for these queries (6 out of 8 cases). Since frequent queries in ORCAS are likely frequent in general, Google suggestions can exploit frequency information from their vast interaction logs (which are absent from ORCAS).

To gain more insights into \sm{the effects of training data frequency}, we qualitatively evaluate examples of generated intents. For instance, the query ``gmat prep classes'' (which only occurs once in ORCAS, as a verbatim match), generates intents such as ``requirements'', ``registration'', and ``training''. Although these are not perfect matches with the Gold intents (companies that offer these courses, practice exams, tips, \smm{similar tests}, and two navigational intents), they are clearly preferable to \craig{the} Google suggestions, which focus on specific locations (e.g., ``near me'', ``online'', ``chicago'', etc.), and demonstrate the ability for the IntenT5 model to generalize. For the query ``used car parts'', which occurs in ORCAS 13 times, IntenT5 generates some of the queries found in ORCAS (e.g., ``near me'') but not others (e.g., ``catalog''). For the query ``toilets'', which occurs 556 times in ORCAS, IntenT5 again generates some queries present in the training data (e.g., ``reviews'') and others that are not (e.g., ``installation cost'').

\looseness -1 These results answer RQ2: IntenT5 effectively generalizes beyond what was seen in the training data. However, it can struggle with cases that occur frequently. This suggests that an ensemble approach may be beneficial, where intents for infrequent queries are generated from IntenT5, and intents for frequent queries are mined from interaction data \sm{(if available). We leave this to future work.}

\subsection{RQ3: Types of Under-specification}\label{ssec:res:rq3}

\begin{table}
\centering
\caption{Diversification performance by query type.  Significant differences between the unmodified system are indicated by~* (paired t-test, Bonferroni correction, $p<0.05$).}
\label{tab:bytype}
\begin{tabular}{lcrrr}
\toprule
&& \multicolumn{3}{c}{$\alpha$-nDCG@20} \\
\cmidrule(lr){3-5}
System         &Agg.& Faceted     & Ambiguous    & Single     \\
\midrule
DPH            &    &   0.3804 &   0.2525 &   0.6399 \\
\greyrule
 + IntenT5     & PM & * 0.4093 &   0.2576 & \bf  0.6711 \\
 + Google Sug. & PM & * 0.4130 &   0.2629 &   0.6621 \\
 + Gold        & PM & \bf * 0.4626 & \bf   0.2908 &   0.6399 \\
\greyrule
 + IntenT5     & xQ &   0.3922 &   0.2433 &   0.6550 \\
 + Google Sug. & xQ & * 0.4070 &   0.2724 & \bf   0.6553 \\
 + Gold        & xQ & \bf * 0.4565 & \bf   0.2997 &   0.6399 \\
\midrule
Vanilla BERT   &    &   0.3809 &   0.2501 &   0.5323 \\
\greyrule
 + IntenT5     & PM & * 0.4244 & \bf   0.3087 & \bf * 0.6415 \\
 + Google Sug. & PM & * 0.4149 &   0.2993 &   0.5353 \\
 + Gold        & PM & \bf * 0.4392 &   0.2831 &   0.5253 \\
\greyrule
 + IntenT5     & xQ & * 0.4214 & \bf * 0.2997 & \bf * 0.6421 \\
 + Google Sug. & xQ &   0.4086 &   0.2970 &   0.5681 \\
 + Gold        & xQ & \bf * 0.4409 &   0.2850 &   0.5253 \\
\midrule
monoT5         &    &   0.4184 &   0.3012 &   0.6172 \\
\greyrule
 + IntenT5     & PM &   0.4445 &   0.3165 & \bf  0.6429 \\
 + Google Sug. & PM &   0.4405 & \bf   0.3342 &   0.6340 \\
 + Gold        & PM & \bf * 0.4802 &   0.3307 &   0.6177 \\
\greyrule
 + IntenT5     & xQ &   0.4363 &   0.3214 &   0.6285 \\
 + Google Sug. & xQ & * 0.4385 & \bf * 0.3327 & \bf   0.6353 \\
 + Gold        & xQ & \bf * 0.4607 &   0.3182 &   0.6177 \\
\midrule
ColBERT        &    &   0.4237 &   0.3267 &   0.5655 \\
\greyrule
 + IntenT5     & PM & * 0.4629 & \bf   0.3246 & \bf * 0.6848 \\
 + Google Sug. & PM &   0.4558 &   0.3198 &   0.6268 \\
 + Gold        & PM & \bf * 0.4740 &   0.3068 &   0.5655 \\
\greyrule
 + IntenT5     & xQ & * 0.4596 & \bf  0.3355 & * 0.6816 \\
 + Google Sug. & xQ & * 0.4536 & \bf  0.3355 &   0.6102 \\
 + Gold        & xQ & \bf * 0.4775 &   0.3016 &   0.5655 \\
\midrule
(Count)         &    & (189)    & (62)     & (49)     \\
\bottomrule
\end{tabular}\vspace{-\baselineskip}
\end{table}

Recall that under-specified queries can be considered multi-faceted or ambiguous. To answer RQ3, we investigate the performance of IntenT5 on different types of queries, as indicated by \craig{the} TREC labels. Note that WT13--14 also include a total of 49 queries that are fully specified (``single'', e.g., ``reviews of les miserables''). Table~\ref{tab:bytype} provides these results. We find that IntenT5 excels at handling faceted queries, often yielding significant gains. When it comes to ambiguous queries, however, IntenT5 rarely significantly improves upon the baseline. Note that all intent strategies, including when using the Gold intents, struggle with ambiguous queries. \io{However, we} acknowledge that the ambiguous query set is rather small (only 62 queries). This could motivate the creation of a larger ambiguous web search ranking evaluation dataset in the future to allow further study of this interesting and challenging problem. Finally, we notice that IntenT5 can also improve the performance of \io{the} fully-specified queries, most notably for Vanilla BERT and ColBERT where the \sm{non-diversified models} otherwise significantly \craig{underperform} DPH. Curiously, we do not observe similar behavior for monoT5, suggesting that this behavior may depend on the underlying language model (BERT vs.\ T5).

\looseness -1 These results answer RQ3: IntenT5 improves the diversity of multi-faceted queries and even improves ColBERT's performance \smm{for} fully-specified queries. However, like alternative approaches, it struggles to generate effective intents for ambiguous queries.

\subsection{RQ4: Handling Ambiguous Queries}\label{ssec:res:rq4}

\begin{table}
\centering
\caption{Diversification performance by query type when using distributional causal language modeling (DCLM) and representation swapping (RS). Statistically significant differences between the unmodified IntenT5 approach are indicated by~* (paired t-test, Bonferroni correction, $p<0.05$).}
\label{tab:dclmrs}
\begin{tabular}{lcrrr}
\toprule
&& \multicolumn{3}{c}{$\alpha$-nDCG@20} \\
\cmidrule(lr){3-5}
System         &Agg.& Faceted     & Ambiguous    & Single     \\
\midrule

monoT5 + IntenT5  & PM &   0.4445 &   0.3165 &   0.6429 \\
 + DCLM           & PM &   0.4424 &   0.3219 & \bf  0.6628 \\
 + RS             & PM &   0.4481 & \bf 0.3279 &   0.6297 \\
 + DCLM + RS      & PM & \bf 0.4457 &   0.3213 &   0.6173 \\
\greyrule
monoT5 + IntenT5  & xQ & \bf  0.4363 &   0.3214 &   0.6285 \\
 + DCLM           & xQ &   0.4333 & \bf   0.3421 & \bf   0.6469 \\
 + RS             & xQ &   0.4341 &   0.3348 &   0.6404 \\
 + DCLM + RS      & xQ &   0.4324 &   0.3129 &   0.6249 \\
\midrule

ColBERT + IntenT5 & PM &   0.4629 &   0.3246 & \bf  0.6848 \\
 + DCLM           & PM & * 0.4339 & \bf   0.3255 &   0.6584 \\
 + RS             & PM & \bf   0.4645 &   0.3185 & \bf   0.6848 \\
 + DCLM + RS      & PM &   0.4420 &   0.3174 &   0.6356 \\
\greyrule
ColBERT + IntenT5 & xQ & \bf  0.4596 & \bf  0.3355 & \bf  0.6816 \\
 + DCLM           & xQ &   0.4469 &   0.3260 &   0.6795 \\
 + RS             & xQ &   0.4564 &   0.3263 & \bf  0.6816 \\
 + DCLM + RS      & xQ &   0.4478 &   0.3250 &   0.6795 \\

\bottomrule
\end{tabular}
\end{table}

Given that ambiguous queries appear to be difficult to handle, we investigate two proposed approaches for overcoming this problem: Distributional Causal Language Modeling (DCLM, introduced in Section~\ref{ssec:meth:dclm}) and Representation Swapping (RS, introduced in Section~\ref{ssec:meth:rs}). Since monoT5 and ColBERT most effectively use IntenT5 on ambiguous queries, we focus our investigation on these models.

Table~\ref{tab:dclmrs} presents the effectiveness of these approaches, stratified by query type. In general, we observe only marginal differences by using combinations of these approaches. The most effective combination for ambiguous queries (monoT5 + IntenT5 + DCLM + xQuAD) is not significantly more effective than the monoT5 + IntenT5 + xQuAD.

Digging deeper into the queries generated for each approach, we find that there are indeed cases where the generated intents using DCLM and RS are substantially more diverse than the base IntenT5 model. The top intents generated for the query \textit{penguins} by  IntenT5 are \textit{meaning}, \textit{history}, \textit{habitat}, \textit{information}, and \textit{definition}; in fact, all of the top 20 intents either relate to the animal (rather than the hockey team) or are very general. Meanwhile, DCLM overcomes many of the general intents, but the queries skew heavily toward the hockey team: \textit{schedule}, \textit{website}, \textit{wikipedia}, \textit{highlights}, and \textit{merchandise}. This problem is addressed when applying both DCLM and RS, which generates: \textit{wikipedia}, \textit{tickets}, \textit{population}, \textit{schedule}, and \textit{website}; it covers both senses. Despite the clear benefits for some queries, the approach can cause drift on other queries, and sometimes does not pick up on important intents. For instance, the intents generated for the query \textit{iron} with IntenT5 + DCLM + RS focus heavily on the nutrient sense, and do not identify the element or appliance sense.

To answer RQ4, although approaches like DCLM and RS can improve the diversity in isolated cases, there is insufficient evidence that these approaches can improve ranking diversity overall. \sm{We also find no significant differences in effectiveness between the DCLM and RS approaches.}

\begin{table}
\centering
\caption{Top intents generated using our IntenT5 model and Google search suggestions.}
\label{tab:examples}
\begin{tabular}{lll}
\toprule
IntenT5 & IntenT5 + DCLM + RS & Google Suggestions \\
\midrule
\multicolumn{3}{l}{\bf penguins} \\
meaning & wikipedia & of madagascar \\
history & tickets & schedule \\
habitat & population & score \\
information & schedule & hockey \\
definition & website & game \\
\midrule
\multicolumn{3}{l}{\bf mitchell college} \\
football & tuition & baseball \\
meaning & football & covid vaccine \\
address & athletics & athletics \\
basketball & admissions & basketball \\
website & bookstore & of business \\
\midrule
\multicolumn{3}{l}{\bf wendelton college} \\
address & tuition & \textit{(none)} \\
football & athletics &  \\
website & bookstore &  \\
tuition & faculty &  \\
application & address &  \\
\midrule
\multicolumn{3}{l}{\bf electoral college} \\
meaning & wikipedia & map \\
definition & meaning & definition \\
florida & definition & map 2020 \\
history & articles & definition government \\
michigan & election & college votes \\
\midrule
\multicolumn{3}{l}{\bf solar panels} \\
meaning & installation & for sale \\
explained & installed & for home \\
calculator & installation cost & cost \\
installation & on sale & for rv \\
home depot & for home & for house \\
\midrule
\multicolumn{3}{l}{\bf condos in florida} \\
for sale & rentals & on the beach \\
meaning & beachfront & for rent \\
near me & for sale & on the beach for sale \\
reviews & near me & keys \\
by owner & reservations & keys for sale \\
\midrule
\multicolumn{3}{l}{\bf condos in new york} \\
meaning & for sale & for rent \\
near me & to rent & zillow \\
chicago & address & manhattan \\
florida & weather & state \\
for sale & nyc & ny \\
\midrule
\end{tabular}
\end{table}

\section{Analysis}\label{sec:analysis}

One advantage of performing search result diversification explicitly is that the generated intents are expressed in natural language and can be interpreted. In Table~\ref{tab:examples}, we present the top 5 intents generated by our models, as well as the top query suggestions from Google. For the running example of \textit{penguins}, we see that Google identifies two senses (an animated film and the hockey team) while our model can identify the animal and the hockey team. For the query \textit{mitchell college}, our model identifies several salient facets, as do the Google search suggestions. Note that this is not due to memorization; the only queries with the text \textit{mitchell college} in the training collection are \textit{william mitchell college of law} and \textit{william mitchell college of law ranking}. This quality is appealing because it shows that the model is capable of generalizing beyond its training data. On the other hand, our model can be prone to constructing information, 
such as for the (fictitious) \textit{wendleton college}. We see that these generalizations are not baked entirely into the prompt of \textit{college}, however, given that the prefix \textit{electoral college} (a process of the United States government) does not generate similar queries.
\sm{These results provide qualitative evidence for our observations in Section~\ref{ssec:res:rq2}; IntenT5 is able to effectively generalize beyond what is seen in the training data.}
However, we acknowledge that this quality may be undesirable in some circumstances. For the query \textit{solar panels}, we see that our model can generate multi-word \sm{intents (which can be beneficial to neural models~\cite{Dai2019DeeperTU})}, but can sometimes get stuck on common prefixes (e.g., ``install''). We also find that our model can struggle with providing valuable recommendations based on a specified location. Although IntenT5 with DCLM and RS can predict salient intents like \textit{beachfront} and \textit{nyc} for the queries \textit{condos in florida} and \textit{condos in new york}, respectively, the base IntenT5 model relies primarily on generic intents, or even suggests alternative locations. Meanwhile, the Google suggestions are able to consistently provide location-specific intents.
\smm{Overall, this analysis shows that IntenT5 generates intents that exhibit awareness of the query at hand, and that the DCLM and RS approaches can change the output of the model substantially. The intents are often comparable with those provided by a commercial search engine \craig{from} interaction data.}

\section{Conclusions}

\looseness -1 \sm{We presented IntenT5, a new approach for generating potential query intents for explicit search result diversification. Across the TREC WebTrack 2009--2014 datasets (300 test queries in total), we found that this approach can significantly outperform other sources of query intents when used with unsupervised search result diversification algorithms and neural re-rankers\smm{, such as Vanilla BERT and ColBERT}. \smm{Specifically, we \io{observed} up to a 15\% relative improvement above query suggestions provided by Google (as measured by NRBP, when re-ranking with Vanilla BERT and aggregating with PM2).} The \io{proposed} approach significantly improves the performance on multi-faceted queries and can even overcome shortcomings on fully-specified queries. We found that IntenT5 has difficulty handling ambiguous queries, and \io{proposed} two approaches for overcoming these ambiguities. Although we observed that these approaches can qualitatively improve the generated intents, we found insufficient evidence that these modifications are beneficial in aggregate. This motivates the creation of a \io{larger and} more extensive dataset of ambiguous queries for \io{a} future study. Importantly, we observed that our approach can generalize to queries that were not present in the training data.} \smm{As the first work to investigate the use of contextualized language models in the context of search result diversification, we \io{ have laid} the groundwork for investigating ongoing challenges, such as handling query terms with multiple senses.}

\begin{acks}
This work has been supported by EPSRC grant EP/R018634/1: Closed-Loop Data Science for Complex, Computationally- \& Data-Intensive Analytics.
\end{acks}

\bibliographystyle{ACM-Reference-Format}\balance
\bibliography{sample-base,macavaney,craig}

\end{document}